\documentclass[12pt]{article}
\pdfoutput=1
\usepackage{cite,graphicx,amsmath,amssymb}
\usepackage{graphicx}
\usepackage{multicol}
\usepackage{xcolor}

\newcommand{\be}{\begin{equation}}
\newcommand{\ee}{\end{equation}}
\newcommand{\bea}{\begin{eqnarray}}
\newcommand{\eea}{\end{eqnarray}}
\begin{document}

\thispagestyle{empty}
\vspace*{-.2cm}
\noindent

\hfill 27 May 2013

\vspace*{1.2cm}

\begin{center}
{\Large\bf AdS/CFT for Accelerator Physics\\
or
\\[.4cm]
Building the Tower of Babel}
\\[2cm]

{\large Arthur Hebecker\\[6mm]}

{\it
Institut f\"ur Theoretische Physik, Universit\"at Heidelberg, 
Philosophenweg 19,\\ D-69120 Heidelberg, Germany\\[3mm]

{\small\tt (\,a.hebecker@thphys.uni-heidelberg.de\,)} }
\\[1.6cm]

{\bf Abstract}
\end{center} 

\noindent
The crucial property of particle colliders is their ability to convert
(e.g. electrical) energy into the mass of heavy particles. We have become 
used to the extremely low efficiency of this conversion and the severe
limitations on the mass scale of heavy particles which can be reached.
In view of this situation, it appears reasonable to ask whether a perfect
conversion machine of this type (a perfect `collider') exists even in 
principle and whether there is a highest mass scale which can be reached 
by such a machine. It turns out that, with a number of assumptions, such 
a machine is conceivable in a world with a strongly-coupled, approximately
scale invariant 4d field theory with 5d gravity dual. This machine can 
be realized as a 5d tower built on the IR brane (in Randall-Sundrum model 
language). Transporting mass to the tip of this tower is, under 
certain conditions, equivalent to producing heavy point-like 4d particles.
Hence, this can be thought of as a perfect `collider'. In the simple, `pure 
Randall-Sundrum setting' that we analyse, this machine can only reach a 
certain maximal energy scale, which falls as the gravity-dual of the 4d QFT 
approaches the strong coupling domain. On these grounds, one might expect 
that a no-go theorem (in the spirit of that of Carnot for the conversion of 
heat into work) exists for generic weakly-coupled QFTs. We end with some 
speculations about collider efficiencies at weak coupling, involving 
possibly the concept of entanglement entropy.

\newpage
\section{Introduction}
The production of very heavy particles is one of the main goals of modern
experimental particle physics. The method of choice is the acceleration 
of beams of charged particles (i.e. the conversion of electrical into kinetic
energy) and their subsequent collision (i.e. the conversion at least a 
small fraction of that kinetic energy into the mass of heavy particles).
While in practice the longevity of these particles has always been very 
limited, the production of stable very heavy states (such as the famous 
WIMP possibly making up dark matter) is most certainly conceivable.

In this context, one naturally encounters the following apparently very
basic and general question: Does there exist, at least in principle, a
perfect machine for the conversion of work into mass of heavy particles?
To give an extreme example, is it conceivable to take just $543\,$kWh 
$=1.22\times 10^{19}$ GeV from the electrical grid and covert them into one 
Planck mass particle?

It is, of course, well known that conventional colliders with this
energy reach are very hard to imagine. Furthermore, even the production of 
e.g. 100 Higgs bosons is energetically much more expensive than the 
equivalent amount of electrical energy would suggest. But the question 
remains whether this is just due to our insufficient ingenuity or the limited
technological progress made so far by mankind, or whether there exists 
some fundamental limitation.

Unfortunately, the present paper fails by a large margin to answer this 
extremely interesting question. However, it will at least outline a somewhat
unusual (AdS/CFT-based \cite{Maldacena:1997re}) way to think about problems 
of this type. In this context, a suggestion for a perfect energy conversion 
machine can be made. It will turn out that the reach of this type of machine 
is limited (at least in the simplest, pure-gravity models to be specified 
below). This range becomes small as the underlying 4d QFT becomes weakly 
coupled. One may interpret this as a hint at the existence of a fundamental 
no-go theorem for a prefect machine in 4d weakly coupled QFT (which is what 
we are apparently stuck with in this part of the multiverse).

The paper is organized as follows: Sect.~\ref{rs} shows that, if our world were 
described by a Randall-Sundrum (RS) model \cite{Randall:1999ee}, a 5d tower 
built on the IR brane (and ideally reaching the UV brane) can be thought of as 
a perfect (Planck scale) collider. This is almost obvious since a simple 5d 
elevator, using electrical energy with an energy conversion efficiency near 
unity, could now be employed to `UV-shift' massive particles. Thus, as a 
first step, a very simple `toy-tower' (a horizontal mirror supported by 
radiation pressure) is considered. It turns out that, at least if one ignores 
all 5d field VEVs except the metric, only a limited height can be 
reached. Making use of specific, non-metric 5d VEVs (as suggested e.g. 
by stringy settings), possibilities for avoiding this maximal heigth 
restriction exist. It remains open whether this loophole can actually be 
turned into a toy-model perfect collider of arbitrary energy reach.

Sect.~\ref{to} discusses an actual tower (made from some imagined 5d analogue 
of solid matter consisting e.g. of 5d `atoms') and its maximal height. The 
result turns out to be similar to that of the previous section (assuming
again only metric VEVs in 5d). Thus, both models suggest that only a certain 
maximal energy can be reached by our ideal collider. The maximal 
height of the tower (and hence the `collider' energy) falls with growing 5d 
curvature. As mentioned above, this conclusion may be avoided in certain 
supersymmetric models with flat directions. Nevertheless, we believe that
our findings suggests that at least in large classes of weakly coupled 4d 
QFTs (where the AdS dual is strongly curved), no `perfect collider' can be 
built even in principle. We attempt to support this by some simple 
considerations concerning the maximal energy conversion efficiency of 
colliders directly in 4d. We will also comment on some of the previous ideas 
concerning `Planck scale colliders', see e.g. \cite{Casher:1995qc,
Labun:2010wf,Banados:2009pr}.

The final section is devoted to a brief summary, a discussion of open 
questions, and further speculations.

\section{Colliders vs. elevators in the Randall-Sundrum model}
\label{rs}

\subsection{How towers in RS models can be used to produce heavy 
particles}

\begin{figure}
 \begin{center} 
  \includegraphics[width=0.3\textwidth]{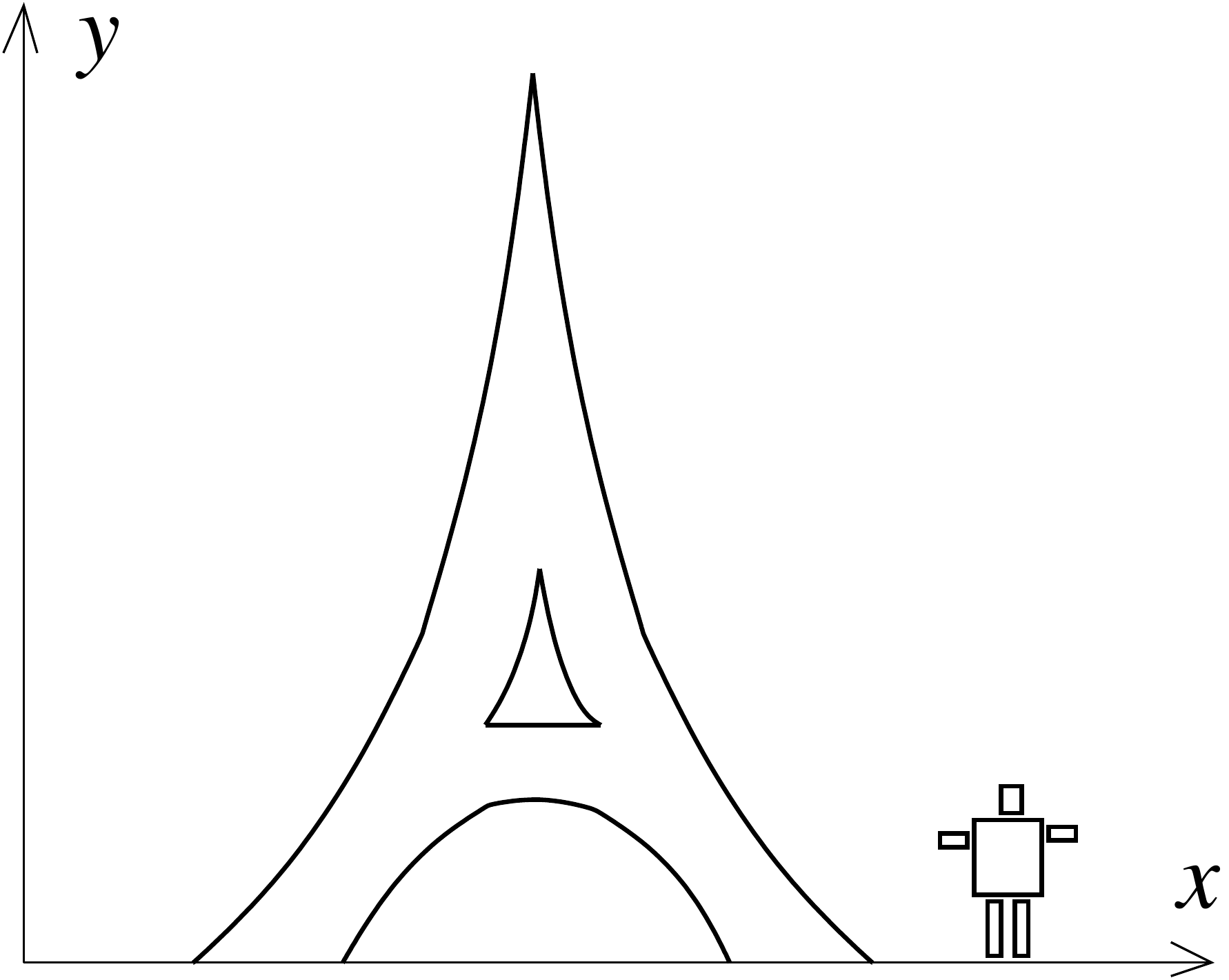} 
 \end{center} 
 \caption{A tower standing on the IR brane of the Randall-Sundrum model, 
built by a `5d robot' which is able to manipulate 5d (i.e. 
sub-TeV$^{-1}$-sized) structures.}\label{tower} 
 \end{figure}

Our use of the AdS/CFT proposal will be limited to its simple yet very 
concrete and intuitive implementation in RS type models. To be very specific, 
we take the AdS metric in the form
\be 
ds^2=e^{2ky}dx^2+dy^2\,,                         \label{metric}
\ee
where $k$ sets the AdS curvature scale. Our discussion is based on the 
action \cite{Randall:1999ee}
\be
S=\int_0^{y_{UV}} d^4xdy\sqrt{-g_5}\left(\frac{1}{2}M_5^2{\cal R}-
{\cal L}_{5d}\right)+\int d^4x\sqrt{-g_{IR}}{\cal L}_{IR}+
\int d^4x\sqrt{-g_{UV}}{\cal L}_{UV}\,.\label{rsa}
\ee
Here the compact space is the interval $y\in [0,y_{UV}]$, with gravity and 
some 5d field theory in the bulk and two 4d theories at the boundaries
(coupled to the induced metrics $g_{IR}$ and $g_{UV}$). Appropriate 
5d and 4d cosmological constants have been absorbed in the 
lagrangians for brevity. As in the celebrated proposal for the solution of 
the hierarchy problem \cite{Randall:1999ee}, we take `our' QFT to be 
IR-brane-localized. Furthermore, and this is the crucial and non-trivial 
step, we imagine that future technology will allow us to penetrate the bulk 
and construct `5d robots' capable of manipulating structures in 5d, at least 
near the IR brane (cf. Fig.~\ref{tower}). 

To be very clear, the point here is not that an RS model will actually be 
discovered at the LHC. Neither do we really hope that we will learn to 
manipulate structures at length scales of TeV$^{-1}$ (which is equivalent to 
manipulating structures in the bulk). We are here considering a `model 
universe', not too dissimilar from our own, where the question of probing the 
Planck scale appears with an interesting twist (as we will presently explain). 

Before doing so, we recall some familiar facts about the setting described 
above (so far without any robots) and its AdS/CFT interpretation (see e.g. 
\cite{Rattazzi:2000hs,Heemskerk:2009pn}): First, we dimensionally reduce to 
4d and Weyl-rescale the 4d metric $g_4$ to ensure that $g_4=g_{IR}$. The 
resulting 4d effective theory of this compactification includes 4d gravity 
(with a Planck scale set by $M_4^2\sim kM_5^3\exp(2ky_{UV})$) and a strongly 
coupled sector (the KK modes of 5d gravity and ${\cal L}_{5d}$). This sector is 
approximately conformal in the energy range $k\ll E \ll k\exp(ky_{UV})$. 
Furthermore, the 4d effective theory also includes the two (by assumption 
weakly coupled) 4d field theories governed by ${\cal L}_{IR}$ and 
${\cal L}_{UV}$. If these two lagrangians, as they appear in (\ref{rsa}), 
are governed by mass parameters $M_1$ and $M_2$, then the two corresponding 
sectors of the resulting 4d effective theory will be governed by mass 
parameters $M_1$ and $M_2\exp(ky_{UV})$ respectively. This is the due to 
the different induced metrics at the two boundaries of our slice of AdS 
space. For simplicity, we set $M_1=M_2=M$ from now on.

From the 4d perspective, this setting looks rather conventional: One
may think of it as of the `Standard Model' (${\cal L}_{IR}$ with mass scale 
$M\sim$ TeV), some form of technicolor, 4d gravity, and a weakly coupled 
sector with very heavy particles (${\cal L}_{UV}$ with mass scale 
$M\exp(ky_{UV})$). The point is that, if we can build a 5d tower (in the 
AdS interpretation of this model, cf. Fig.~\ref{tower}), then this 
corresponds to a perfect collider (in the sense of a 
machine for producing very heavy point-like particles)
on the 4d side. We will shortly estimate the maximal height our 5d tower can 
reach, but before doing so let us argue in some detail that such a tower 
would be able to do the job of a conventional particle collider: Indeed, let 
us assume that ${\cal L}_{5d}$ contains some fundamental field of mass $m$ 
($m \sim M$ for simplicity). Corresponding particles can hence be produced by 
a conventional (i.e. IR-brane-bound) collider. This 5d field may also couple 
to a set of UV-brane fields, allowing e.g. its decay to two UV-brane 
particles of mass $\epsilon m$ and $(1-\epsilon)m$. Thus, if a 5d tower 
reaching the UV brane could be build, this would be equivalent to a perfect 
Planck scale collider: One would just have to create our 5d particles with a 
TeV-scale machine, transport them up the tower using conventional mechanical 
energy (e.g. in an elevator) and eventually let them decay to UV-brane 
particles of mass almost equal to $m$. From a 4d perspective, this 
corresponds to producing heavy, point-like particles (since ${\cal L}_{UV}$ 
is supposed to be a weakly-coupled local lagrangian) of mass 
$m\exp(ky_{UV})$ with energy conversion efficiency $\eta_{coll.} \sim 1$. (Here 
we ignore the (in)efficiency of our original 4d collider taking us up to the 
TeV domain.)

\subsection{Toy model of a suspended mirror}\label{toy}

\begin{figure}
 \begin{center} 
  \includegraphics[width=0.3\textwidth]{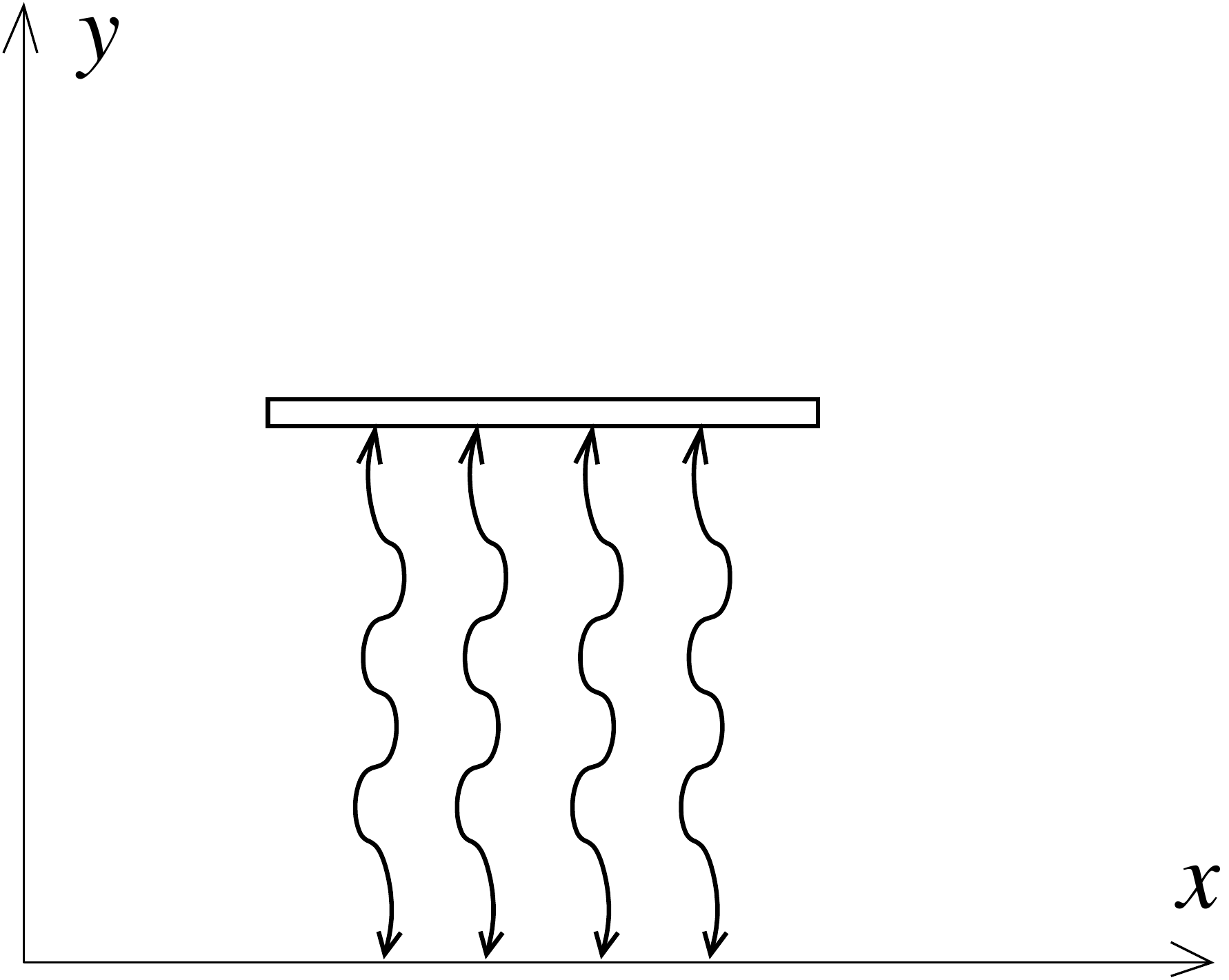} 
 \end{center} 
 \caption{Mirror supported by a `5d photon' beam above the IR 
brane.}\label{mirror} 
 \end{figure}

Now it is unfortunately clear that a tower of some particular desired height 
(e.g. reaching the UV brane) can not be built in general. To understand the 
limitations, let us first focus on an (at least calculationally) simpler 
device which is sufficient for suspending an elevator: We add 5d photons to 
our list of assumptions and let a mirror float above the IR brane, 
supported by the pressure of photons bouncing back and forth between brane 
and mirror (cf. Fig.~\ref{mirror}). Obviously, to construct the mirror and 
the elevator, we also have to assume that some form of structured, stable 
matter exists in 5d.\footnote{
This 
is non-trivial since all structures we manipulate every day in 4d rely 
microscopically on renormalizable gauge theories, which are not available 
in 5d. In particular, it is well-known that the Schr\"odinger atom is 
unstable if $d>4$ (see e.g. \cite{Tangherlini:1963bw}). Let us nevertheless 
assume that some form of structured matter can exist at $d=5$ (e.g. because 
a full QFT treatment cures the non-relativistic instability problem) and 
press ahead.
}
Governed by our 4d experience, we take this matter to consist of some small
units (`atoms'). To simplify our analysis, we assume these `atoms' to have 
mass and inverse size $M$.\footnote{
Obviously,
our familiar 4d atoms have a mass and size which are parametrically different 
since the former is governed by the mass of the nucleus while the latter 
depends on electron mass and gauge coupling. In this language, our 5d model 
of matter corresponds to taking $m_N\sim m_e$ and $\alpha_e\sim 1$ in 4d.
}
We are interested in the lightest possible mirror, which will nevertheless 
have a thickness of at least a few `atoms'. The (hyper)surface density of this
object will hence be $\rho_s\sim M^4$. (Note that this mirror extends in 
3 spatial dimensions and hence the corresponding surface density has units 
of mass/(length)$^3$.) 

To determine the force required to support such a mirror, consider first a 
particle with mass $m$ that is stationary at some height $y$. We assume 
that no non-gravitational field VEVs are present or are at least not 
relevent in the present context. (This assumption will be removed in 
Sect.~\ref{bf}.) The relevant action is
\be
S_y=-m\int_{y=\mbox{\scriptsize const.}} d\tau = -m\,e^{ky}\int dt\,,
\ee
where $\tau$ and $t$ are the eigentime and the time at the IR brane 
respectively. It is apparent that the same particle, if stationary at height 
$y+\delta y$, has an action enhanced by a factor $\exp(k\,\delta y)$. Thus, 
`lifting' a particle a distance $\delta y$ costs an energy 
\be 
\delta E = m\,e^{k(y+\delta y)}- m\,e^{ky}\simeq m\,e^{ky}\,k\,\delta y\,
\ee
from the perspective of the IR brane. Here the factor $e^{ky}$ appears as 
a `blue-shift', because we took the IR-brane point of view. For a local
observer at height $y$, lifting the same particle by $\delta y$ costs an 
energy $\delta E \simeq m\,k\,\delta y$. The force required to support 
a particle $m$, and now we use the local perspective, is hence $km$. 

Our mirror is supported by the vertically directed (both up and down) photon 
stream with energy momentum tensor
\be
T_{MN} \sim \mbox{diag}(\rho,p,p,p,p) = \mbox{diag}(\rho,0,0,0,\rho)\,,\qquad
\mbox{where}\qquad M,N\in \{ 0,1,2,3,5 \} \,,
\ee
which is here given in a coordinate system with Minkowski metric in the 
vicinity of the mirror. To keep the mirror stationary, we need
\be
p = \frac{F}{A} = \frac{\rho_s A k}{A} =\rho_s k \sim M^4 k\,,
\ee
in self-explanatory notation. This is the pressure (and hence energy density)
at the position $y$ of the mirror. Since each photon travels vertically (at 
constant $\vec{x}$), the number of photons per unit-brane-surface (in the 
vicinity of the IR brane) is enhanced by $\exp(3ky)$. Furthermore, due to the 
gravitational redshift, each photon has an energy enhanced by $\exp(ky)$ when 
it is reflected by the IR brane. Thus, the energy density of our beam 
near the brane is 
\be
\rho_{IR} \sim M^4 k e^{4ky}\,.                   \label{rhoir}
\ee

Assuming that the reflection of photons both at the IR brane and at our 
mirror is perfect, we can imagine that this configuration is stationary,
without the need of continuous energy input. Nevertheless, the mirror had
to be raised to its position $y$, which required the input of energy into 
the photon beam near the IR brane. Since we assume that such an energy
input can be realized maximally at a scale $M$, we have the constraint 
$\rho_{IR}< M^5$. Comparing this with (\ref{rhoir}), we see that the maximal
height $y_{max}$ which can be achieved is set by
\be
e^{ky_{max}} \sim \left(\frac{M}{k}\right)^{1/4}\,.  \label{ekym}
\ee

In fact, there is an additional constraint arising from the danger of 
black hole formation (or, more generally, strong deformation of the 5d 
metric) in the region of high beam density. To see this, note that we
actually have a layer of thickness $\sim 1/k$ of an approximate 
energy density $\rho_{IR}$ directly above the IR brane. We now estimate how 
large $\rho_{IR}$ can become before black holes are formed in this region. 
To do so, recall that the mass of a $d$-dimensional black hole of radius $R$ 
is (see e.g. \cite{Tangherlini:1963bw,Myers:1986un})
\be
M_{BH} \sim M_{P,d}^{d-2}R^{d-3}\,.        \label{bh}
\ee
This has to be compared to the relation between mass and radius of the 
corresponding smooth energy distribution:
\be
M_{BH} \sim R^{d-1}\rho\,.                          \label{smooth}
\ee
Eliminating $M_{BH}$ from (\ref{bh}) and (\ref{smooth}) and specifying 
$d=5$, we determine the critical radius for black hole formation, 
\be
R_c\sim \sqrt{\frac{M_5^3}{\rho}}\,,  \label{rc}
\ee
where $M_5$ is the 5d Planck mass. Now we substitute $R_c\sim 1/k$ and 
$\rho\equiv \rho_{IR}$ (cf. (\ref{rhoir})), in (\ref{rc}) and solve for 
$\exp(ky)$. This gives us another bound on the achievable height $y$,
\be
e^{ky}\sim\frac{M_5^{3/4}k^{1/4}}{M}\,,   \label{bhb}
\ee
supplementing (\ref{ekym}). 

One possible interpretation is that (\ref{ekym}) remains our basic formula
for the maximal height but, due to (\ref{bhb}), we in addition need to 
demand 
\be
M_5^3>\frac{M^5}{k^2}\,,                  \label{m5c}
\ee
i.e., 5d gravity has to be sufficiently weak. As outlined earlier, we assume 
that our mirror has been lifted together with an attached 5d elevator, such 
that we are now in possession of a collider with `energy reach' $(M/k)^{1/4}$. 
In other words, we can use e.g. photons at energy $M$ to produce particles 
with mass $M(M/k)^{1/4}$, with $100\%$ energy efficiency (at least in 
principle). Obviously, we here do not include the one-time energy investment 
required for the construction of this `collider'. 

For example, the UV brane or `Planck brane' of \cite{Randall:1999ee} could 
be located at the height $y_{max}$ given by (\ref{ekym}), in which case 
we could `lift' energy to the Planck brane. Note that, due to the constraint 
(\ref{m5c}), the 4d Planck mass ($M_4^2\sim M_5^3/k$ using the UV-brane induced
metric) remains higher than $M$, such that we can never actually reach the 
4d Planck scale using this type of `perfect collider'. 

It is obvious that our construction with a horizontal mirror and a vertical 
photon beam is far from optimal. It can be improved by making the floating 
mirror as small (in brane-parallel direction) as possible, curving it 
appropriately, and supporting it by a tapering photon beam arrangement. This 
clearly requires an appropriate mirror array at the IR brane. We do not 
pursue this analysis here but turn, in Sect. \ref{to}, to the construction of 
an (also tapering) `real' tower made from solid material.

\subsection{Including bulk fields beyond the metric}\label{bf}
A natural objection is that, in `proper' string-theoretic AdS/CFT 
\cite{Maldacena:1997re}, a D3-brane can, due to the BPS condition, be 
stationary at any point in the radial direction of AdS$_5$. Thus, in models 
with such BPS objects, it appears to be easy to avoid the height restrictions 
found in the last subsection.

For simplicity, we implement the key ingredients directly in our 5d RS 
setting: Let us assume that the 5d action contains a 5-form field strength 
$F_5=dC_4$ and 3-branes charged under $C_4$. Furthermore, let us supplement 
the gravitational background of (\ref{metric}) by an $F_5$-VEV proportional 
to the volume form. For an appropriately tuned value of this field strength 
(our RS-model analogue of the type IIB BPS condition), such a brane can rest, 
in parallel to the IR brane, at any value of $y$: The gravitional force is 
precisely compensated by the force of the field strength permeating the bulk. 

Before continuing, we note that, from the perspective of `generic' strongly 
coupled 4d models with 5d gravity dual, the above are rather special 
requirements: If no such, infinitely extended, brane is present in a given 
4d vacuum, it can not be created by any means (unlike a mirror, which can be 
assembled from `atoms'). Second, if no 5-form VEV is present or it is not 
approriately tuned, it is impossible for any potential experimentalist to 
create one. Equally, the charge of the 3-brane can not be adjusted, as the 
charge of the electron can not be adjusted in our 4d world. 

In principle, one may consider theories in which 3-branes are allowed to 
have boundaries (this clearly requires further types of charges and gauge 
fields etc., but let us assume this can be engineered). Such a finite brane 
could then be created, but it would not be BPS: It has a finite tension which 
will in general lead to the shrinking of its area. A detailed technical 
analysis of whether one could stabilize such a brane and what the 
energetic cost of `lifting' this stabilized configuration would be goes 
beyond the goals of this paper.

\begin{figure}
 \begin{center} 
  \includegraphics[width=0.8\textwidth]{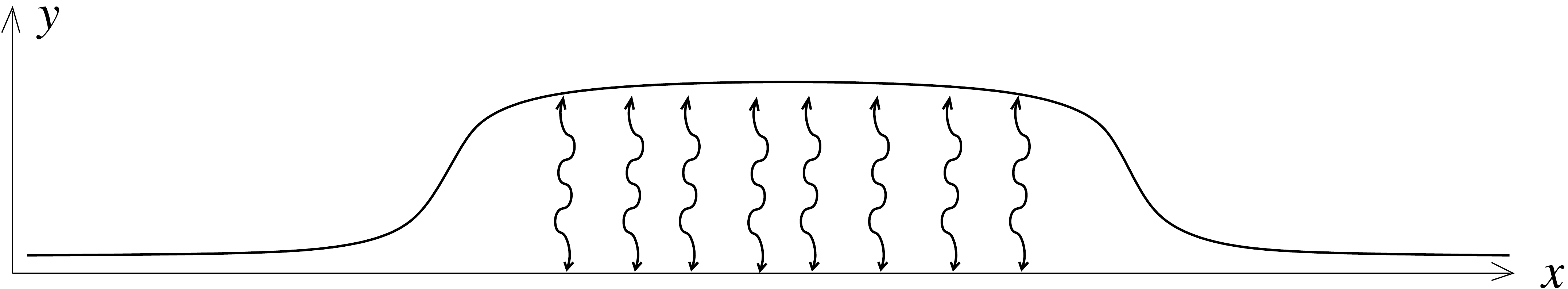} 
 \end{center} 
 \caption{Infinitely extended 3-brane lifted in a finite spatial region. 
Note that the `photon beam' in the figure is only meant to symbolize some 
mechanism for displacing the brane. It is unlcear whether such a fundamental 
3-brane can play the role of an actual mirror for some form of radiation.}
\label{3-brane} 
\end{figure}

Let us instead continue with the arguably more `natural' case of an 
infinitely extended BPS 3-brane, assuming one was present. Indeed, one might 
imagine `lifting' such a brane in the (necessarily finite) spatial region 
accessible to an experimentalist (cf. Fig.~\ref{3-brane}). In 4d language, 
this amounts to living in a vacuum with a modulus (the 3-brane position 
in $y$-direction) and shifting this modulus within a finite spatial domain. 
It might in principle be interesting to follow this route and see whether 
one can attach some structure (presumably non-BPS) to this `partially lifted' 
brane. We do not want to pursue this in the present paper since we already 
have a clear understanding how this setting avoids the proposed `no go 
theorem' in 4d language: One needs a modulus, which can then be displaced 
through some `experimental effort'. Clearly, this allows one to explore 
totally new physics. Whether it can lead to the Planck scale and to a 
reversible process for transforming macroscopic work to the mass of Planck 
scale particles remains open at the moment.

Inspired by the above, one may however consider other bulk fields with VEVs
in analogy to the 5-form field strength naturally suggested by type IIB 
string theory. Insisting on 4d Poincare symmetry, the only options are a 
4-form or, equivalently, the Hodge-dual 1-form field strength. The latter 
is just the gradient of a scalar, which anyway has to be 
present in a complete model because of Goldberger-Wise stabilization 
\cite{Goldberger:1999uk}. One is now allowed particles (clearly simpler 
than a 3-brane) coupling to this background scalar field. Thus, one may hope 
to compensate the gravitational force on those particles through this 
coupling. 

Let us be slightly more specific by considering a 5d fermion $\psi$,
\be
{\cal L}_5 \supset -i\overline{\psi}\gamma^M D_M \psi -\overline{\psi}\psi 
f(\phi)\,,
\ee
where $f$ is some function of the Goldberger-Wise scalar $\phi$. In analogy 
to the discussion at the beginning of Sect.~\ref{toy}, one can convince 
oneself that the 4d energy of such a $\psi$-particle (at rest in $x^\mu$) is
\be
E_\psi(y) = m_5(y)e^{ky}=f(\phi(y)) e^{ky}\,.
\ee
Assuming that the scalar background $\phi(y)$ is a monotonic function, we can 
without loss of generality (through a field redefinition $\phi \to \phi'(
\phi))$) work with the specific background $\phi(y)=cy$. If we furthermore 
take $f(\phi)=\exp(-k\phi/c)$, we obviously obtain $E_\psi(y)=\mbox{const}$. 
As a result, $\psi$ particles can be easily moved to the Planck brane, in 
analogy to the BPS 3-brane discussed earlier. Equally obviously, however, a 
machine doing this is {\it not} a perfect collider since the particles do not 
become extremely massive (in 4d language) when moved to the UV. 

Nevertheless, it is interesting to consider the 4d analgue of the above 
situation: In 4d, the $\psi$-particles are composites characterized my some 
mass $m_4$ and size $l_4$. Given the right choice of $f$, they possess a flat a 
direction (the analogue of the $y$-position of the elementary 5d particle 
$\psi$) on which $l_4$ (but not $m_4$) depends. As $y$ increases, $l_4$ 
becomes tiny, making the particle point-like in 4d language. To ensure that 
this particle in addition has a Planck-scale mass, $m_4$ would need to be 
$\sim M_4$ from the start. In other words, we need a theory with very heavy, 
extended objects (e.g. some type of soliton) which can be `manufactured' using 
macroscopic work. These objects would also need to possess a flat direction 
along which their size changes. Clearly, this is rather exotic (even more so 
than the rest of the paper) and we choose not to pursue this line of thinking 
for now.

\section{Maximal-height 5d towers and possible implications for
4d colliders at weak coupling}\label{to}

\subsection{Optimal towers}
We return to the simplest possible setting without any non-metric VEVs in 5d.
An optimal tower will use the strongest 5d material available, i.e. that with
the largest ratio $p/\rho$.\footnote{
Presumably 
$p/\rho\ll 1$ holds even for the strongest available material, at least if 
this material is made from point-like weakly-interacting particles, as in our 
4d world. Note, however, that our world is not weakly-coupled throughout
and that much stronger materials, such as the neutron star crust, appear to 
exist \cite{Horowitz:2009ya}. 
}
We will henceforth assume that this ratio is maximized for one particular 
substance, which we will use to build our tower. Most naively, one would
try to adapt Weisskopf's famous argument \cite{Weisskopf:1970jx} for the 
maximal height of mountains (expressed in terms of fundamental constants) 
to our situation. While his argument is energetic (sinking of the mountain 
vs. melting of the rock at the bottom of the mountain), we make an 
essentially equivalent force-based estimate:

First, as a warm up, let $ky\ll 1$ such that $\exp(ky)\simeq 1+ky$. A
rectangular 5d mountain with (constant!) cross section $A$ and height $y$
has mass $Ay\rho$ and exerts a force $Ay\rho k$ on its base. The base 
can provide a force $Ap$. Hence, for a given constant $p/\rho$ the maximal 
height is
\be
y_{max}=\frac{1}{k}\cdot\frac{p}{\rho}\,.
\ee
While self-consistent with our linearization (since $p/\rho<1$), this is 
clearly not interesting: The crucial energy reach of our `collider'
is $\exp(ky_{max})$, which can hence not become large in our toy-model 
with constant cross section. 

An optimal tower will taper towards its tip, such that each cross section is 
just large enough (assuming maximal vertical pressure at each point of the 
cross section) to support the part of the tower above. This clearly can be 
cast in the form of a differential equation for the cross section $A(y)$, 
and we will do so shortly. The solution then determines the shape of the 
tower and, as we will see, its maximal height. 

Naively, one might expect to find a complete solution of this simple and 
fundamental problem in engineering textbooks or papers. However, in 
real-world towers, wind pressure is the most important issue and (unlike our 
case) the gravitational field can be treated either as linear 
($\exp(ky)\to gy$) or, if one considers extremely high towers, according to 
the $1/r^2$ force-law. The closest related ideas and calculations in the 
literature appear to be related to either the Tsiolkovsky tower or the space 
elevator \cite{Tsiolkowskii} suspended from a point in geostationary orbit 
(in the latter case, the tapering is towards the bottom for obvious reasons). 
In any case, we were not able to find a treatment of a situation exactly 
equivalent to ours. 

Fortunately, the corresponding equations are simple even in our exotic case.
Everything can be derived from an equation relating the vertical forces at 
heights $y$ and $y+\delta y$:
\be
F(y)=F(y+\delta y)\cdot (1+k\delta y) + k \rho A(y) \delta y\,.   \label{force}
\ee
Except for the factor $(1+k\delta y)$, this is self-evident: Going down the 
tower by a distance $\delta y$, the force grows by the weight of an additional 
layer of material. The factor $(1+k\delta y)$ comes from the warping: As
explained earlier, raising a mass from $y$ to $y+\delta y$ costs an energy 
$mk\delta y \exp(ky)$ from the perspective of $y=0$. This means that this 
mass exerts a force $mk\exp(ky)$ at any support at $y=0$, while it 
obviously only exerts a force $mk$ at any support at its own height. In other
words, vertical forces are subject to warping in the very same way as 
energies. Thus, the weight of all the tower material above $y+\delta y$ 
exerts a force on the surface at height $y$ which is enhanced by a factor 
$\exp(k\delta y)\simeq (1+k\delta y)$. This is the content of the first
term on the r.h. side of (\ref{force}).

With $F(y)=p A(y)$ and $p=\mbox{const.}$ (an optimal tower will have maximal 
pressure at any layer), one then immediately derives a differential equation 
for $A$,
\be
-A'(y)=A(y)k(1+\rho/p)\,,
\ee
where $\rho$ is constant by assumption. The solution is 
\be
A(y)=A_0 e^{-(1+\rho/p)ky}\,.                \label{ay}
\ee
Just to prevent any possible confusion: As should be clear from the 
derivation, this function $A(y)$ characterizes the $y$-dependence of 
the cross-section of our tower as a locally well-defined 5d physical 
quantity. For example, it could be the cross section in 5d Planck 
units. It is very different from the cross section as measured in the 
coordinates $x^\mu$ of (\ref{metric}). 

In our analysis of the shape of the tower we have neglected any horizontal
force components. This is only justified as long as the tower is a `thin 
object', i.e., $A(y)$ does not change too rapidly with $y$. Quantitatively, 
this will certainly hold if the angle between the tower surface and 
the vertical axis is small. Most naively, one would estimate this angle 
as (minus) the derivative of the tower radius with respect to the height: 
$-[A^{1/3}(y)]'$. However, due to warping this derivative is non-zero even for 
a vertical tower, i.e. for a tower the surface of which is made from lines 
at $\vec{x}=\mbox{const.}$ In fact, the cross section of such a vertical 
tower is given by $A_v(y)=A_0\exp(3ky)$. Thus, when estimating the angle 
at the base of the tower and requiring it to be parametrically small, we 
have to do so relative to vertical tower:
\be
-\Big\{\,[A^{1/3}(y)]'-[A_v^{1/3}(y)]'\Big\}=\Big\{\frac{(1+\rho/p)k}{3}+k
\Big\}A_0 \ll 1\,.
\ee
This translates into an estimate of the maximal $A_0$ allowed:
\be
A_0^{1/3} \sim \frac{3}{(4+\rho/p)k}\,.              \label{a0}
\ee

At its tip, our tower can certainly not become thinner than $1/M$. Thus, 
substituting  $A_0^{1/3}$ from (\ref{a0}) and $A(y)^{1/3}\sim 1/M$ in 
(\ref{ay}), we eventually find that the maximal height $y_{max}$ is determined 
by
\be
e^{ky_{max}}\sim \left(\frac{3M}{(4+\rho/p)k}\right)^{\frac{3}{1+\rho/p}}\,.
\ee
Note that this is rather similar to our `floating mirror' result of 
(\ref{ekym}): Since we did not keep track of ${\cal O}(1)$ factors, the 
prefactor $3/(4+\rho/p)$ accompanying the ratio $M/k$ is most probably
irrelevant. The only difference is then in the exponent. For an 
isotropic 5d radiation gas, which is presumably close to the 
stiffest possible matter, we have $p=\rho/4$ and hence an exponent 
$3/5$. This is better than the $1/4$ of (\ref{ekym}), although we have 
to remember that we did not try to optimize the shape of the beam in 
Sect.~\ref{rs}. Thus, the competition between the two `perfect collider
technologies' of Sects.~\ref{rs} and \ref{to} can not be decided at this
level of precision. 

It is interesting to note that the approximate agreement arises in spite of 
the two configurations being distinctly different: The tower we are 
presently constructing becomes wider towards its base. By contrast, the 
region of the IR brane from which the photon beam of Sect.~\ref{rs} is 
reflected is much smaller than the floating mirror. 

Finally, we expect a bound on $M_5$ arising from the danger of black hole
formation at the base of the tower. It is easy to obtain by requiring that 
the critical radius of (\ref{rc}) is smaller than the width of the tower
at its base, given by (\ref{a0}). One finds
\be 
M_5^3>\frac{9\rho}{(4+\rho/p)^2k^2}\,,               \label{m5cc}
\ee
which, for the natural value $\rho\sim M^5$, once again becomes extremely 
similar to the analogous bound of (\ref{m5c}).

\subsection{Strong coupling appears to be unavoidable}\label{scu}

To sum up, we have seen in two independent ways that a perfect collider 
with energy reach $(M/k)^\alpha$ (with $\alpha\sim {\cal O}(1)$) can be 
built in principle. Since we are using a weak-coupling analysis on the 
gravity side, the corresponding 4d theory has to be strongly coupled. Let 
us try to be more precise by recalling that \cite{Maldacena:1997re}
\be
\lambda \sim g_{YM}^2 N \sim \left(\frac{M_s}{k}\right)^4
\ee
in `proper' AdS/CFT, i.e. in the duality between 4d ${\cal N}=4$ 
Super-Yang-Mills theory and type IIB string theory on $AdS_5\times S^5$. 
Here $\lambda$ is the `t Hooft coupling, i.e. the actual control parameter 
of perturbation theory on the 4d side, and $M_s\sim 1/l_s$ is the string 
scale. Thus, if we were to identify our `scale of 5d structure' $M$ with 
the string scale $M_s$, we would find that the energy reach of our collider 
grows as $\lambda \to \infty$. By contrast, it approaches unity for 
$\lambda \to 1$. In other words, our perfect collider exists precisely
because the 4d theory is strongly coupled and it ceases to exist as we
approach the boundary between strong and weak coupling. Clearly, this 
does not exclude the existence of perfect colliders of some totally 
different type in the weak coupling domain, but it is a hint to the 
contrary.

We continue the discussion of a possible stringy realization of our 5d 
collider by recalling that, dimensionally reducing from 10d to 5d on $S^5$, 
we have 
\be
M_5^3\sim M_s^8/k^5\,.
\ee
This is consistent with the constraint (\ref{m5c}) or, equivalently, 
(\ref{m5cc}) if one identifies $M$ with $M_s$ as suggested above. 
Unfortunately, there is a different problem with this identification:
At the scale $M_s$, we are deeply in the 10d regime since the 
compactification scale from 10d to 5d is $k$. Thus, we would either have
to redo our `tower building exercise' in 10d or construct the tower on 
branes extending along the radial direction of the $AdS_5\times S^5$ throat. 
Furthermore, it may be necessary to use a `structure scale' below $M_s$ to 
avoid back-reaction. We leave the investigation of these interesting 
questions to future work.

We end this line of thought be recalling that, to introduce an IR scale, all 
of the above would actually have to be done in a Klebanov-Strassler throat 
\cite{Klebanov:2000hb} (or one of its variants) rather than in the 
$AdS_5\times S^5$ throat. The former has a well defined IR end (an `IR brane', 
if one wishes) and can be consistently glued to a Calabi-Yau\footnote{
For 
a different view on fundamental limitations in learning about the internal 
geometry of string compactifications see e.g. \cite{Heckman:2013kza}.
}
at its UV end, thereby inducing 4d gravity. It's length is automatically 
stabilized, which can be viewed as a variant of Goldberger-Wise stabilization 
\cite{Goldberger:1999uk,Brummer:2005sh}. Thus, including the embedding of 
branes, it might be possible to realize all of the above in almost realistic 
settings. The trouble is only that, up to now, the LHC gives us very little 
hope that we actually live in such a geometry with low string scale. Also, 
while it would be interesting to pursue the line of thought of 
Sect.~\ref{bf} in the string-theoretic context, we note that D3-branes are 
not BPS and can not freely float `between IR und UV' in a compactified and 
stabilized model \cite{Kachru:2003sx}.

\subsection{Towards the weak coupling regime}\label{towards}
Given that our world is apparently weakly-coupled, it is most interesting 
to consider implications for the weak coupling regime. Of course, realistic 
accelerator technology is a highly-developed field of research (see e.g. 
\cite{Ankenbrandt:1999cta}) and the present author is completely ignorant in 
this important field. Hence this subsection only serves to share some 
vague ideas inspired by the previous `fundamental' perspective. Three
such ideas or potential relations between the strong-coupling 
and weak-coupling situation are detailed below. Before proceeding, we note 
that an analysis related in spirit to this subsection has appeared 
in \cite{Casher:1995qc}. In contrast to \cite{Casher:1995qc}, our emphasis 
is on the efficiency of energy conversion rather than on fundamental 
achievability of Planck scale energies (which we take for granted by linear 
collider technology, at least in a flat universe and with $M_4\to\infty$). 

\begin{figure}
 \begin{center} 
  \includegraphics[width=0.3\textwidth]{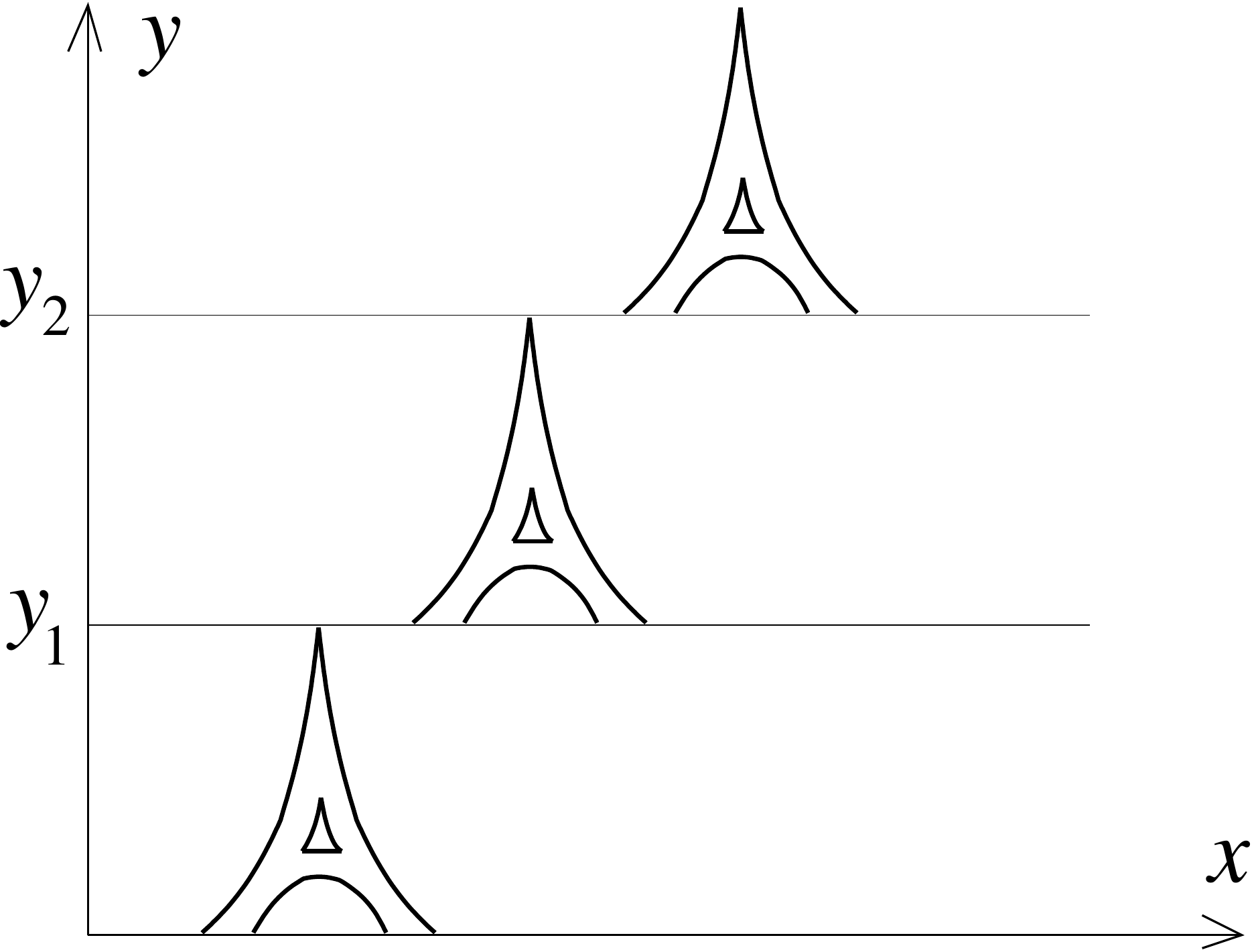} 
 \end{center} 
 \caption{If additional branes in between the IR and UV brane exist, a
cascade of towers can, in principle, reach arbitrarily high energy 
scales.}\label{cascade} 
 \end{figure}

Our first point is related to the possibility that, in between the IR and UV
brane, other branes \cite{Oda:1999di} (stabilized by the Goldberger-Wise 
mechanism \cite{Goldberger:1999uk} at certain 5d positions $y_i$) may exist. 
Given appropriate 4d field theories on those branes and 
couplings between these 4d theories and the 5d bulk fields, 
each brane may serve as a `base' for the construction of a 
tower, cf. Fig.~\ref{cascade}. In other words, if the first brane is low 
enough to be reachable by a tower, a second tower can be constructed starting 
from this brane. This tower could, in principle, reach yet another brane, 
and so on. To summarize, it is conceivable that in such a generalized RS 
model an arbitrarily high energy scale can be reached by a perfect collider.

This peculiar construction has a surprisingly obvious weakly-coupled 4d
analogue. Indeed, imagine the LHC finds stable TeV-scale charged particles 
and some future, purpose-built collider produces them copiously. It is then 
conceivable to fill a further collider (storage ring) with those particles 
and accelerate them (very efficiently because of their huge mass and hence
small synchrotron radiation) to a much higher energy.\footnote{
In 
analogy to the muon-collider, just with much heavier and, most importantly, 
stable muons.
}
If, at this (say PeV) energy, another yet heavier species of stable charged 
particles is found and can be copiously produced, one may go on and reach 
the Planck scale (or at least a very high energy scale). While this 
collider cascade would presumably not be perfect (i.e. $\eta_{coll.}\ll 1$
parametrically), it would nevertheless be much better than anything we can 
imagine without new stable particles. It appears that the collider cascade 
works for the same reason as the tower cascade described above: If 
the fundamental theory is {\it not} approximately conformal, reaching high
energies appears to be more doable, both at weak and strong coupling. 

The second connection with 4d weakly-coupled theories is more straightforward:
Let us ignore all technical objections and envision an extremely long 
(built in open space, far away from stars or planets\footnote{
We 
also ignore the gravitational effects emphasized in \cite{Casher:1995qc}.
})
linear collider reaching a very high energy scale $M_{UV}$. In principle, the 
energetic efficiency of the accelerator part can be close to unity. 
However, the production of heavy particles (i.e. 
the collider part) presumably works much less efficiently for the following 
reason: The quality of beam-focusing may be limited {\it in principle}
(we will return to this in a moment). Let us assume that there is some 
minimal area $1/m^2$ to which the beams can at best be focused near the 
interaction point. The cross section for the production of particles of mass 
$M_{UV}$ is $\sim 1/M_{UV}^2$. Since, due to electromagnetic interactions, the 
beams are lost after the interaction point, the efficiency with which energy 
is converted into mass of heavy particles can not be better than 
\be
\eta_{coll.} \sim m^2/M_{UV}^2\,.\label{etaw}
\ee

Now, the crucial issue is the value of the `optimal focusing scale' $m$.
Even if we assume that transverse oscillations of the beam particles can 
in principle be completely removed by cooling, it is hard to imagine how 
the focusing scale can ever exceed the mass of those particles. This 
expectation follows simply from the uncertainty principle and the limitations 
on the electromagnetic field focusing the beam. (Note that for our 
purposes we identify electron and proton mass and consider the electromagnetic 
coupling as an ${\cal O}(1)$ parameter.) Thus, we identify $m$ in (\ref{etaw}) 
with the `smallest scale at which we can manipulate structures' (e.g. electron 
mass etc.). In other words, $\eta_{coll.}$ falls below unity to the 
extent that $M_{UV}$ exceeds this `structure scale'. But this is exactly what 
the holographic point of view discussed earlier had suggested: In that case, 
the structure scale of the full theory is the IR-brane structure scale 
multiplied by the warp factor of the maximal-height tower. This is the mass 
scale that could be reached by our ideal `collider'. Anything beyond can 
only be reached by sacrificing efficiency. 

According to the above, setups with point-like heavy particles 
bear some resemblence to systems with negative temperature 
\cite{Ramsey:1956zz}. Indeed, the transformation of work into thermal 
energy of a negative-temperature system can be extremely inefficient. This 
fundamental inefficiency is apparently shared by the transformation of work 
into mass of heavy particles. To be more specific, let us assume that our 
heavy particles correspond to energy with negative temperature $T_2$. 
Consider a Carnot cycle with input heat $Q_1$ at positive temperature $T_1$ 
(e.g. $\sim 10^3 K$, as in power stations), output `heat' $Q_2$ at $T_2$, and 
output work $W$. Its efficiency,
\be
\eta = \frac{W}{Q_1}=1 - \frac{T_2}{T_1}\,,
\ee
can be greater than unity. This is intuitive since the negative-temperature 
system can deliver energy while increasing its own entropy (see e.g. 
\cite{rapp}). It can thus contribute {\it positively} to the delivered work, 
$Q_2<0$, while respecting reversibility (no entropy change, $\delta S=0$). 
Our previous collider efficiency was that for the transformation of (input) 
work $W$ to `heat' at $T_2$, corresponding to the reversed process. Hence
\be
\eta_{coll.} = \frac{-Q_2}{W} = \frac{W-Q_1}{W} = 1-\frac{1}{\eta}=
\frac{-T_2}{T_1-T_2}\,.
\ee
This quantity can thus be extremely small as a matter of principle (as 
conjectured in this paper) if the `equivalent' temperature of heavy 
point-like particles is negative and tiny.

Finally, the third connection between the strongly- and weakly coupled 
situations goes as follows: Let us to try imagine, at least very roughly, what 
a direct 4d weak-coupling analogue of our 5d-tower would look like. For 
example, one could think of a spherical mirror inside which a standing wave 
of electromagnetic energy was trapped. The energy density in this spherical
standing wave would have to grow to extreme values as one approaches the 
center. Let alone the question of how to create a sufficiently perfect mirror 
and to set up the relevant field configuration, one immediately sees that 
it could never be stable: The reason is simply that in the inner part 
(where the energy density would have to be extremely high) 
electron-positron pairs would be created and escape to the 
outside. This is just a result of the non-vanishing interaction between UV and 
IR modes, which is unavoidable in a weakly-coupled local field theory. By
contrast, in our tower the highly-concentrated (in 4d language) energy 
at the tip of the tower is prevented from decaying to IR-brane degrees of 
freedom by 5d locality. Thus, we here see another hint that the construction 
of perfect colliders is presumably even harder at weak than at strong 
4d coupling.

\section{Conclusions}
We have presented some, admittedly rather speculative, ideas concerning 
the (im)pos\-sibility of a perfect collider. Our main technical point was 
very simple: For theories having a 5d gravity dual, reaching for UV energy 
scales corresponds to building 5d towers based on the IR brane and pointing
to the UV brane. In many theories, the height of such towers appears to be 
limited at a rather fundamental level (quite analogously to the limited 
height of mountains, given the limited strength of granite). We estimated 
this maximal height and conjectured (given the parametric behaviour of our 
result) that at least in generic 4d weakly-coupled theories it is completely 
impossible to build a perfect machine (i.e. a machine with energetic 
efficiency near unity) which transforms energy, starting from the `structure 
scale' of our theory, towards the UV. 

Clearly, in our holographic approach, accelerator physicists are `tower 
builders', struggling with the 5d gravitational potential. The 4d 
weak-coupling analogue of their problem is apparent: The tendency of 
massive objects to fall translates into the tendency of energy to transfer 
from the UV to the IR in conventional QFT.

In specific models, perfect or near-perfect `colliders' might nevertheless 
exist. Two classes of potential examples are discussed in Sect.~\ref{bf}: 
These are theories with completely flat directions (moduli) in field space 
and theories with solitonic objects possessing such flat directions. The  
naive intuition about the overwhelming force of 5d gravity fails in such 
settings and it is conceivable that a perfect machine transforming energy 
from IR to UV can be build. Another class of examples is described at the 
beginning of Sect.~\ref{towards}: If scale invariance in the energy regime 
between TeV and Plack scale is strongly broken by many intermediate branes 
(RS model perspective) or by many stable charged particle species with masses 
spread throughout this domain (weakly-coupled 4d perspective), reaching the 
Planck scale is at least much easier. Thus, we can not expect a fundamental 
no-go theorem which is completely general. The assumptions of such a 
possible theorem have to include details of the relevant model (i.e. of the 
concrete 4d QFT). Obviously, counterexamples which are as close to the real 
world as possible would be much more exciting than the proof of a no-go 
theorem. 

Many interesting questions are still open: For example, it has to 
be clarified whether a tower is really the only or at least the best way to 
transfer energy reversibly (with efficiency near unity) from the IR brane to 
an arbitrarily high position above it. It is tempting to speculate that the 
presently very popular concept of holographic entanglement entropy 
(see e.g. \cite{Ryu:2006bv,Takayanagi:2012kg,Nozaki:2013wia}) has something 
to do with this: Obviously, by transferring energy to the UV we concentrate 
it in a small 4d area, with a low entanglement entropy. This is nicely 
visualized as concentrating the energy inside the minimal surface measuring 
the entanglement according to \cite{Ryu:2006bv}.

Also independently of the holographic interpretation, entanglement entropy 
may be relevant. For example, an extreme growth of entanglement entropy in 
particle decay (taking the perspective of one of the decay products) has been 
obtained in the analysis of \cite{Lello:2013bva}. But particle decay is just 
the inverse of particle production in a collider. Thus, one may hope that the 
difficulties in constructing (even in a Gedankenexperiment) a perfect collider 
can be understood, at the fundamental level, as follows: Such a collider 
creates very energetic states with near-zero entanglement entropy (the very 
heavy, point-like particles to be produced). This may be impossible with 
high energetic efficiency due to limitations analogous to those familiar from 
the Carnot cycle and its standard entropy-based analysis.\footnote{
Of 
course, this is far from obvious since our source of energy is conventional 
work with entropy zero. If the gravitational IR attraction in the AdS 
slice could be given a clear thermodynamic meaning, e.g. along the lines of 
\cite{Jacobson:1995ab}, a more direct application of the 2nd law of 
thermodynamics may become possible.
} 
Thus, an `entanglement entropy analysis' of an ideal machine transforming 
(macroscopic, entropy-free) work into heavy point-like particles may be 
worthwhile. Alternatively, as we discussed in some detail, a more direct
Carnot-type obstruction to perfect colliders arises if one can argue that
settings with heavy point-like particles correspond to negative-temperature 
systems.

Finally, returning to the holographic perspective, the following 
consideration appears interesting: Envision a universe with structure, 
planets, live etc., but with a relatively large cosmological constant.
If the de-Sitter curvature is high enough, it may prevent the construction 
of a very long linear accelerator as a matter of principle (see 
\cite{Casher:1995qc} for related arguments). Let the field theory in addition 
be strongly coupled, with a gravity dual of the type of a Randall-Sundrum 
model. In that situation also the construction of sufficiently high 5d tower 
(as detailed in the paper) may be impossible. As a result, reaching the UV 
or Planck brane may be ruled out altogether. This could be a fundamental 
obstruction to unraveling microscopic details of 4d quantum gravity - a 
`UV protection' mechanism reminiscent of \cite{Dvali:2010bf}, but different 
at the technical level.\footnote{
One 
may envision producing a large black hole and waiting for it to evaporate, 
thus probing quantum gravity through the observation of the final moments 
of its evaporation. However, the mass inside the horizon may be too small
to produce a black hole or the universe may not live long enough for 
complete evaporation.
} 
Thus, a world in which the UV completion of its field theory is 
absolutely protected from observation (and hence not part of physical 
reality) may be conceivable. We find it intriguing to think that this 
form of `UV protection' is due to the overwhelming force of gravity in the 
dual AdS geometry.

\section*{Acknowledgments}

I would like to thank Stefan Theisen and Timo Weigand for helpful 
discussions.

\end{document}